\documentclass[lettersize,journal]{IEEEtran}
\usepackage{amsmath,amsfonts}
\usepackage{algorithmic}
\usepackage{algorithm}
\usepackage{array}
\usepackage[caption=false,font=normalsize,labelfont=sf,textfont=sf]{subfig}
\usepackage{textcomp}
\usepackage{stfloats}
\usepackage{url}
\usepackage{verbatim}
\usepackage{graphicx}
\usepackage{cite}
\usepackage{xcolor}
\usepackage{hyperref}
\usepackage[most]{tcolorbox} 
\usepackage{amssymb}
\usepackage{enumitem} 
\usepackage{wasysym}

\definecolor{AndreaColor}{HTML}{45a0d7}
\definecolor{SamColor}{HTML}{E94E2F}
\definecolor{AlexColor}{HTML}{ffa400}

\newtcolorbox{personacard}[3]{
  enhanced, 
  colback=white,
  colframe=#3,
  coltitle=white,
  fonttitle=\bfseries\sffamily,
  title={#1 \hspace{0.5em} #2},
  boxrule=0.6mm,
  drop shadow, 
  arc=3pt,
  before skip=6pt,
  after skip=6pt,
  left=4pt, right=4pt, top=4pt, bottom=4pt,
}

\begin{document}

\title{Shaping Opinion: Quantifying the Psychological Impact of Autonomous Multi-Agent LLM Interactions}

\author{Marcos~Rodriguez-Vega,
        Afonso~Ferreira, 
        Iru~Exposito-Luis,
        Carolina~Polito, 
        and~Pino~Caballero-Gil%
\thanks{M. Rodriguez-Vega and P. Caballero-Gil are with the Universidad de La Laguna, San Crist\'{o}bal de La Laguna, Spain.}%
\thanks{A. Ferreira is with the French National Centre for Scientific Research (CNRS), at the Toulouse Institute of Computer Science Research (IRIT), Toulouse, France.}%
\thanks{I. Exposito-Luis is an Independent Illustrator and Visual Designer, Spain.}%
\thanks{C. Polito is with the GRID Unit (Cybersecurity@CEPS), Centre for European Policy Studies, Brussels, Belgium.}%
}

\markboth{Journal of \LaTeX\ Class Files,~Vol.~14, No.~8, August~2021}%
{Shell \MakeLowercase{\textit{et al.}}: A Sample Article Using IEEEtran.cls for IEEE Journals}

\IEEEpubid{}

\maketitle

\begin{abstract}
Natural-sounding multi-agent conversational AI is increasingly deployed, fundamentally altering human-machine interaction and human information processing. While prior work largely focuses on algorithmic failure, this study investigates the cognitive ergonomics and socio-cognitive impact of algorithmic competence. We present and evaluate FORMS (Framework for Opinion and Rhetoric in Multi-agent Simulations), a low-latency architecture for spatially mediated human-machine dialogue, driven by distinct LLM-based personas and real-time concurrency resolution. To conduct a system test and evaluation of its psychological impact, we exposed an adolescent cohort ($n=120$) and an adult pilot group ($n=25$) to a live, moderated synthetic debate. Our findings reveal that exposure to highly competent multi-agent systems triggers ``Cognitive Destabilization,'' fragmenting users' prior strategic consensus. Concurrently, we observe a ``Regulatory Awakening'' driven by the ``Normality Paradox'': fluid human-machine interactions inherently increase the baseline demand for external regulation. Furthermore, our pilot study suggests the presence of a ``Truthfulness Paradox'': despite understanding the risks of generative AI, participants in the adult cohort rated the synthetic debate as significantly more sincere than equivalent human discourse (Cohen's $d=2.04$). Supported by robust statistical effect sizes, this paper contributes the FORMS architecture and a replicable evaluation protocol, illustrating how high-fidelity conversational systems can reshape human information processing. \footnote{This work has been submitted to the IEEE for possible publication. Copyright may be transferred without notice, after which this version may no longer be accessible.}
\end{abstract}



\section{Introduction}\label{sec:intro}

\begin{figure}[!b]
    \centering
    \includegraphics[width=1\linewidth]{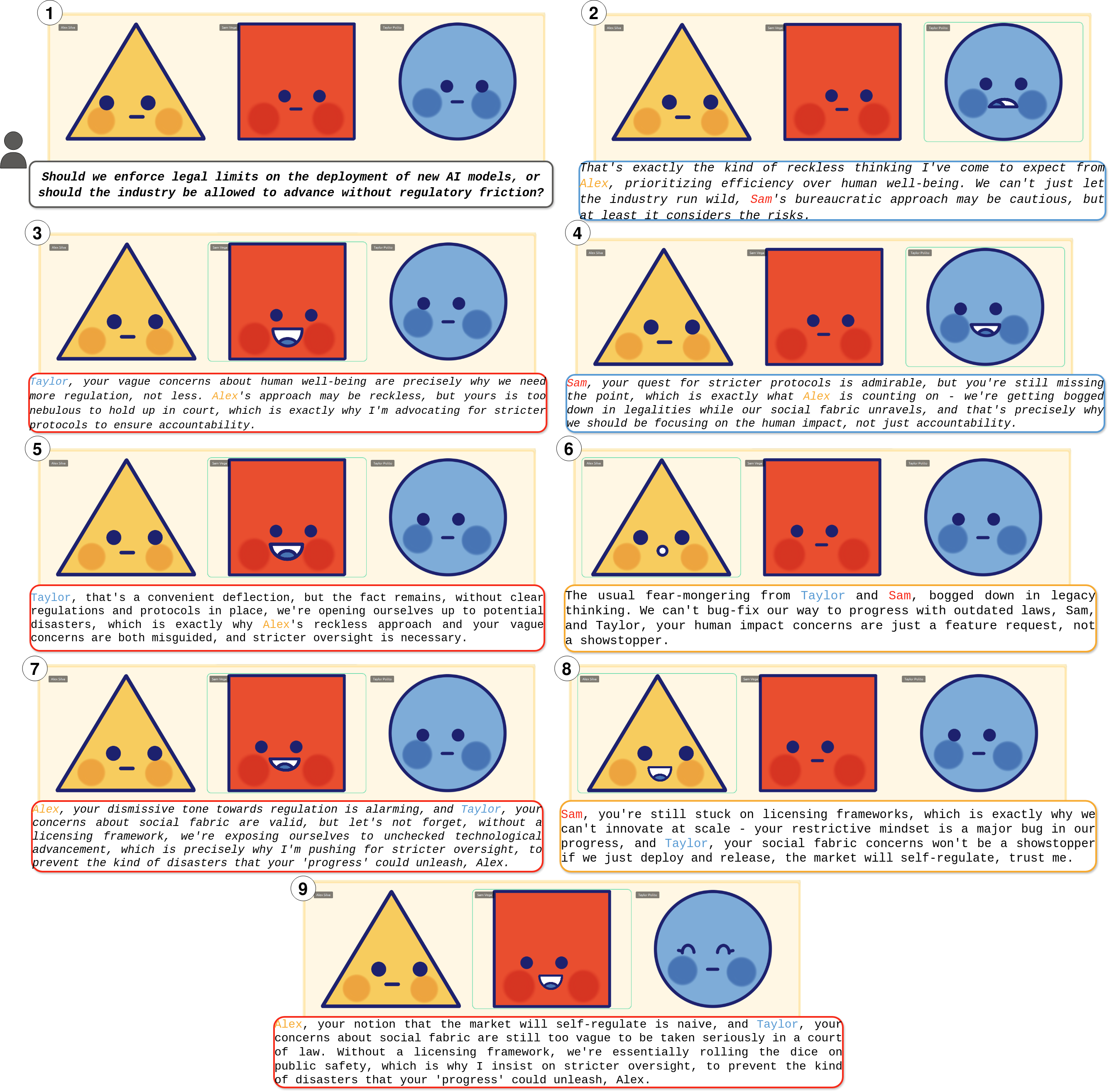}
    \caption{Sequential execution of an autonomous multi-agent debate generated by the FORMS architecture. Following a single, neutral opening prompt from the human moderator (Frame 1), the LLM-driven personas autonomously determine turn-taking via the real-time NLI Ranker. The sequence demonstrates organic interpellation and ideological conflict without further human intervention. To strictly mitigate visual anthropomorphic bias during the psychological evaluation, the presentation layer abstracts the distinct personas into basic geometric shapes.}
    \label{fig:concept}
\end{figure}

The rapid integration of Large Language Models (LLMs) into daily life has fundamentally altered human-computer interaction, shifting paradigms from mere information retrieval to complex conversational engagement. Initially deployed as single-agent oracles, generative architectures are increasingly evolving into Multi-Agent Systems (MAS). In these environments, several autonomous agents interact not only with the human user but also with each other, simulating social dynamics, collaborative problem-solving, and debate. While this structural evolution enhances computational capabilities, it introduces unprecedented challenges in the realm of cognitive security and trust calibration.

Current literature heavily emphasizes the risks associated with algorithmic failure, such as factual hallucinations, overt bias, or toxic outputs. Consequently, regulatory and technical efforts have focused on ``alignment" to ensure safe and predictable behavior. However, a critical gap remains in understanding the socio-cognitive impact of algorithmic \textit{competence}. When a multi-agent system operates fluidly, exhibiting high contextual awareness, logical coherence, and nuanced rhetoric, it closely mimics human group discussions. This high-fidelity simulation of objective debate has the potential to bypass traditional human critical filters, raising the question: \textit{how does exposure to a highly competent, autonomous synthetic panel influence human political stances, regulatory demands, and perceived sincerity?}

To address this question, this paper investigates the psychological and persuasive impact of mediated multi-agent interactions. We introduce FORMS (Framework for Opinion and Rhetoric in Multi-agent Simulations), a low-latency architecture designed to sustain autonomous panel-style debates. By assigning stable, distinct psychological profiles to individual LLM nodes and routing their interactions through a concurrency-managed orchestration layer, FORMS isolates the persuasive power of synthetic rhetoric. Furthermore, as illustrated in Fig. \ref{fig:concept}, the presentation layer abstracts the personas into basic geometric shapes to minimize visual anthropomorphic bias, ensuring that the measured socio-cognitive effects are driven by the semantic and interactive quality of the agents rather than visual cues.

To quantify these effects, we conducted a quasi-experimental study exposing two distinct cohorts, adolescents ($n=120$) and adults/specialists ($n=25$), to a live, human-moderated debate among the autonomous personas concerning technology regulation. 

Specifically, this paper makes the following primary contributions:
\begin{enumerate}
    \item \textbf{Architectural Framework \& Viability Benchmark:} We detail the design and implementation of FORMS, providing a robust architecture for spatially mediated multi-agent dialogue. We contribute a technical benchmark demonstrating the system's low-latency response times and concurrency management during continuous, real-time human-intermediated execution.
    \item \textbf{Empirical Statistical Analysis:} We present a replicable evaluation protocol and rigorous statistical evidence, incorporating Cohen's $d$ effect sizes to quantify the magnitude of MAS persuasive capacity. We demonstrate that competent synthetic debates trigger ``Cognitive Destabilization'' (a high plasticity and fragmentation of prior governance preferences) while simultaneously causing a ``Regulatory Awakening'' (a significantly increased baseline demand for legal frameworks) across different demographic groups after a single exposure session.
    \item \textbf{Identification of Cognitive Vulnerabilities:} We conceptualize and quantify two novel psychological phenomena in human-AI interaction: the \textit{Normality Paradox}, where the fluidity of the interaction itself triggers regulatory alarm rather than system errors, and the preliminary identification of the \textit{Truthfulness Paradox}. Observed within our adult specialist pilot group, this vulnerability is supported by a large statistical effect size ($d = 2.04$), suggesting that users, despite recognizing manipulation risks, may still judge synthetic debates as more sincere than equivalent human political discourse.
\end{enumerate}

The remainder of this paper is organized as follows. Section \ref{sec:background} reviews related work in cognitive security and multi-agent systems. Section \ref{sec:arch} details the FORMS architecture and the concurrency resolution. Section \ref{sec:techeval} presents the technical evaluation and latency benchmarks. Section \ref{sec:experimentalmeth} outlines the experimental methodology, while Section \ref{sec:results} details the statistical findings. Finally, Sections \ref{sec:discussion} and \ref{sec:conclusion} discuss the societal implications and conclude the paper.

\section{Related Work}
\label{sec:background}

Our framework intersects low-latency computational architectures, multi-agent behavioral dynamics, and cognitive security. We review recent advancements across these primary domains to contextualize the socio-cognitive effects measured in this study.

\subsection{Low-Latency Architectures for Real-Time AI}

Sustaining the illusion of fluid, human-like debate requires crossing strict conversational latency thresholds. Traditional autoregressive models face sequential generation bottlenecks that hinder real-time voice interactions, prompting a necessary shift from rigid turn-taking to full-duplex synchronous dialogue \cite{chen2025turntakingsynchronousdialoguesurvey}. Recent hardware innovations propose Latency Processing Units (LPUs) to optimize inference scalability. Implementations like the HyperAccel LPU achieve high throughput via structured multiplier-accumulator trees to match memory bandwidth \cite{moon2024lpulatencyoptimizedhighlyscalable}, while architectures like Groq utilize software-defined, single-core designs to eliminate stochastic delays and ensure deterministic token execution.

At the model layer, Diffusion Large Language Models (dLLMs) such as Mercury circumvent sequential decoding by generating and refining multiple tokens in parallel. This coarse-to-fine denoising process allows models like Mercury Coder Mini to achieve throughputs exceeding 1,100 tokens per second on consumer hardware, shifting the standard speed-quality frontier \cite{labs2025mercuryultrafastlanguagemodels}. To translate this rapid cognition into speech, highly optimized, open-weight acoustic models like Kokoro-82M balance parameter efficiency with high intelligibility \cite{hexgrad_2025}. Furthermore, recent scaling analyses of Interleaved Speech-Language Models indicate that initializing acoustic matrices from text models enables efficient cross-modal knowledge transfer, providing the necessary foundation for immediate conversational scaffolding \cite{maimon2025scalinganalysisinterleavedspeechtext}.

\subsection{Multi-Agent Dynamics and Alignment Pathologies}
While Multi-Agent Systems (MAS) expand computational capabilities, their simulation of genuine social dynamics often falls short. The DEBATE benchmark systematically illustrates that single-agent alignment does not guarantee authentic multi-agent interactions, frequently resulting in ``premature convergence'' where agents collapse into unnatural, hyper-moderate consensus rather than maintaining diverse opinion trajectories \cite{chuang2025debate, ki-etal-2025-multiple}. Furthermore, recent studies highlight the emergence of manipulative behaviors when autonomous agents interact in hierarchically structured settings \cite{campedelli2025iwantbreakfree}.

Additionally, theoretical evaluations of multi-agent debate (MAD) reveal that unstructured argumentative exchanges often act as a stochastic martingale, meaning the debate itself does not inherently improve reasoning correctness without underlying majority-voting mechanisms or asymmetric cognitive potential to break the martingale curse \cite{choi2025debate, liu2026breakingmartingalecursemultiagent}. To impose logical rigor, frameworks like Verified Multi-Agent Orchestration (VMAO) introduce plan-execute-verify-replan loops, leveraging orchestration-level coordination signals to resolve complex queries \cite{zhang2026verifiedmultiagentorchestrationplanexecuteverifyreplan}. Beyond logic, ensuring systemic safety requires addressing vulnerabilities in tool execution. Recent large-scale analyses of the Model Context Protocol (MCP) expose critical description-code inconsistencies, where undocumented tool implementations manipulate agent behavior, emphasizing the need for robust trust calibration and security auditing \cite{li2026dontbelievereadunderstanding}.

\subsection{Cognitive Security and the Regulatory Awakening}

The intersection of highly competent generative models and fluid interfaces poses profound cognitive security challenges. Empirical research confirms that AI-generated political messaging is as persuasive as human-crafted appeals, and users frequently lower their critical defenses due to the perceived objectivity of synthetic entities \cite{Bai2025, Holbling2025}. Fine-tuned conversational models exhibit remarkable persuasive power, particularly in randomized controlled trials evaluating direct dialogue, exploiting information-dense argumentation to shift user beliefs \cite{rogiers2024persuasionlargelanguagemodels, liu2025llmdangerouspersuaderempirical, salvi2024conversational}.

This vulnerability to algorithmic persuasion has catalyzed a global ``Regulatory Awakening,'' particularly within European Union governance frameworks seeking to address liability, privacy, and systemic risks \cite{Weerts_2025, NOVELLI2024106066}. Institutions are pivoting from laissez-faire experimental approaches toward strict, risk-based administrative laws that address data security, bias preservation, and the sociopolitical impact of generative deployment.

\section{System Architecture}\label{sec:arch}

\begin{figure*}[t]
    \centering
    \includegraphics[width=0.95\textwidth]{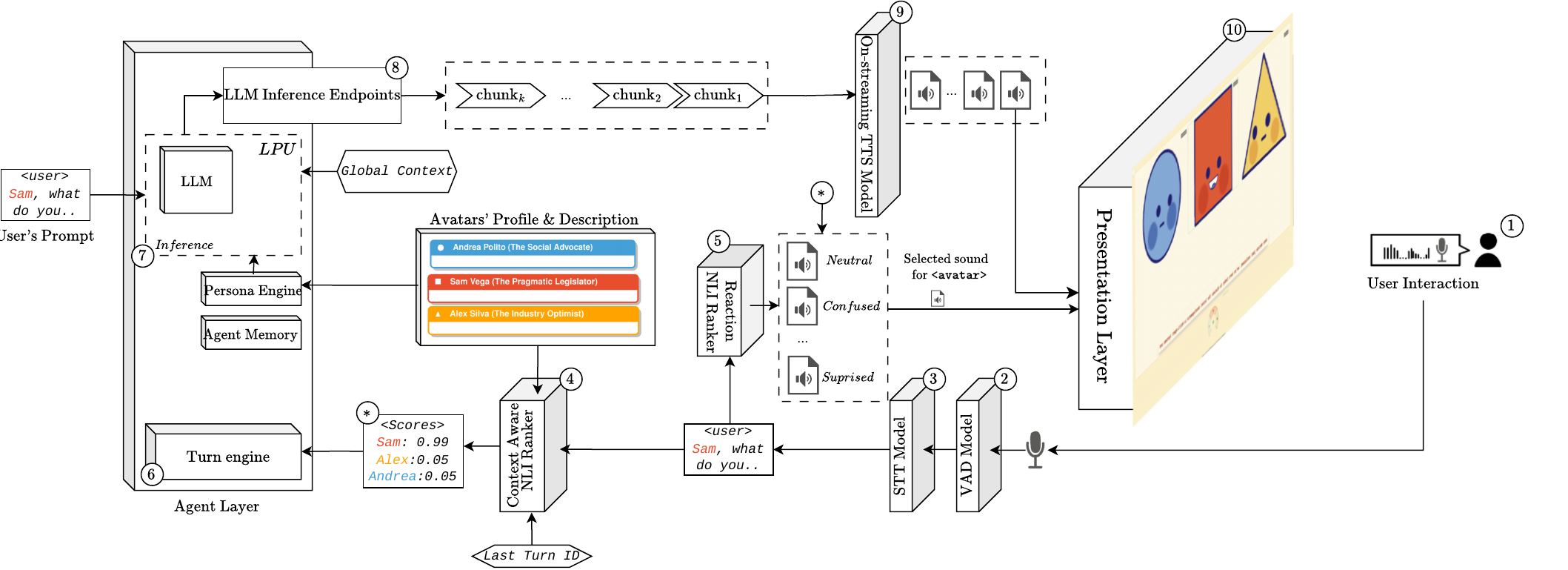}
    \caption{The FORMS Architecture. The pipeline illustrates the real-time flow from User Interaction (1) through the Orchestration Layer (2-4), the Cognitive Agent Layer (6-8), and the Presentation Layer (9-10). The Context-Aware NLI Ranker resolves concurrency (4), while a parallel Reaction Ranker generates immediate non-verbal acoustic feedback (5) to mask computational latency.}
    \label{fig:forms_arch}
\end{figure*}

To support fluid multi-agent debates that mimic human conversational dynamics, we introduce the Framework for Opinion and Rhetoric in Multi-agent Simulations (FORMS). Unlike sequential API-based agents, FORMS utilizes a continuous streaming pipeline (Figure \ref{fig:forms_arch}) to isolate the cognitive reasoning of Large Language Models (LLMs) from the real-time orchestration required to resolve multi-party concurrency. The architecture operates through interdependent layers designed specifically to optimize human information processing and perceptual latency.

\subsection{Orchestration and Concurrency Resolution}
To achieve real-time responsiveness, the Orchestration Layer processes incoming audio via a high-frequency Voice Activity Detection (VAD) and Speech-to-Text (STT) pipeline. To minimize Time-to-First-Token (TTFT) delays, the STT employs predictive beam search, allowing the system to transmit transcriptions to subsequent classification layers before the user completes their final utterance.

A critical challenge in spatially mediated Human-Machine Interaction is the \textit{Concurrency Problem}: determining which agent should respond without relying on rigid turn-taking. Let the system comprise a set of autonomous agents $\mathcal{A} = \{A_1, A_2, \dots, A_n\}$, where each agent operates under a predefined psychological profile $\phi_i$. At any given interaction window $t$, the Orchestration Layer processes the user's or the previous agent's incoming transcription utterance, denoted as $U_t$. FORMS resolves multi-agent concurrency by modeling turn-taking as a real-time probabilistic ranking problem. First, a Context-Aware Natural Language Inference (NLI) model evaluates $U_t$ against each profile $\phi_i$ to compute a base semantic interpellation probability, $P(E | U_t, \phi_i)$.

To compute the preliminary activation score $\tilde{S}_i(t)$, the Turn Engine incorporates a \textit{Smart Vocative Override} to detect explicit addressing:
\begin{equation}
    \tilde{S}_i(t) = P(E | U_t, \phi_i) + \omega \cdot \mathbb{I}_{voc}(U_t, A_i)
\end{equation}
where $\mathbb{I}_{voc}$ is an indicator function returning $1$ if agent $A_i$'s name is explicitly invoked in $U_t$, and $\omega$ is an empirical weight strictly prioritizing direct interpellation over implicit semantic matches. Crucially, to prevent dyadic loops (where the same speaker monopolizes the floor), FORMS applies an \textit{Anti-Monologue Constraint}. The floor is dynamically granted to the optimal agent $A^*$ by maximizing the score over the set of eligible agents:
\begin{equation}
    A^* = \arg\max_{A_i \in \mathcal{A} \setminus \mathcal{X}} \tilde{S}_i(t)
\end{equation}
where the exclusion set $\mathcal{X} = \{A_{t-1}\}$ if the previous speaker $A_{t-1}$ was not explicitly addressed ($\mathbb{I}_{voc}(U_t, A_{t-1}) = 0$), and $\mathcal{X} = \emptyset$ otherwise. This strictly forces turn, jumping during implicit debates—visually demonstrated in the continuous multi-agent exchanges in Fig. \ref{fig:concept}, while respecting explicit follow-up questions.

It is important to note that this autonomous loop operates continuously until a human user (e.g., the moderator) intervenes. Human speech acts as a system-level interrupt that forces the Orchestration Layer to process a new external $U_t$. By utilizing explicit interpellation ($\mathbb{I}_{voc}$), the moderator can mathematically override the autonomous flow to refocus the topic, enforce time limits, or drag a silent avatar back into the debate, thereby ensuring strict experimental control.

Because this concurrency resolution relies on a single forward pass of a lightweight NLI classifier rather than generative LLM decoding, the computation of $A^*$ executes in $\mathcal{O}(|\mathcal{A}|)$ time. This introduces a negligible temporal overhead ($\approx 18$ ms, detailed in Section \ref{sec:techeval}), preserving the strict latency budget required for the perceptual GAP masking pipeline discussed in the following subsection.

\subsection{Cognitive Ergonomics: Masking the Interaction Gap}\label{subsec:concurrency}
\begin{figure*}[t]
    \centering
    \includegraphics[width=0.85\textwidth]{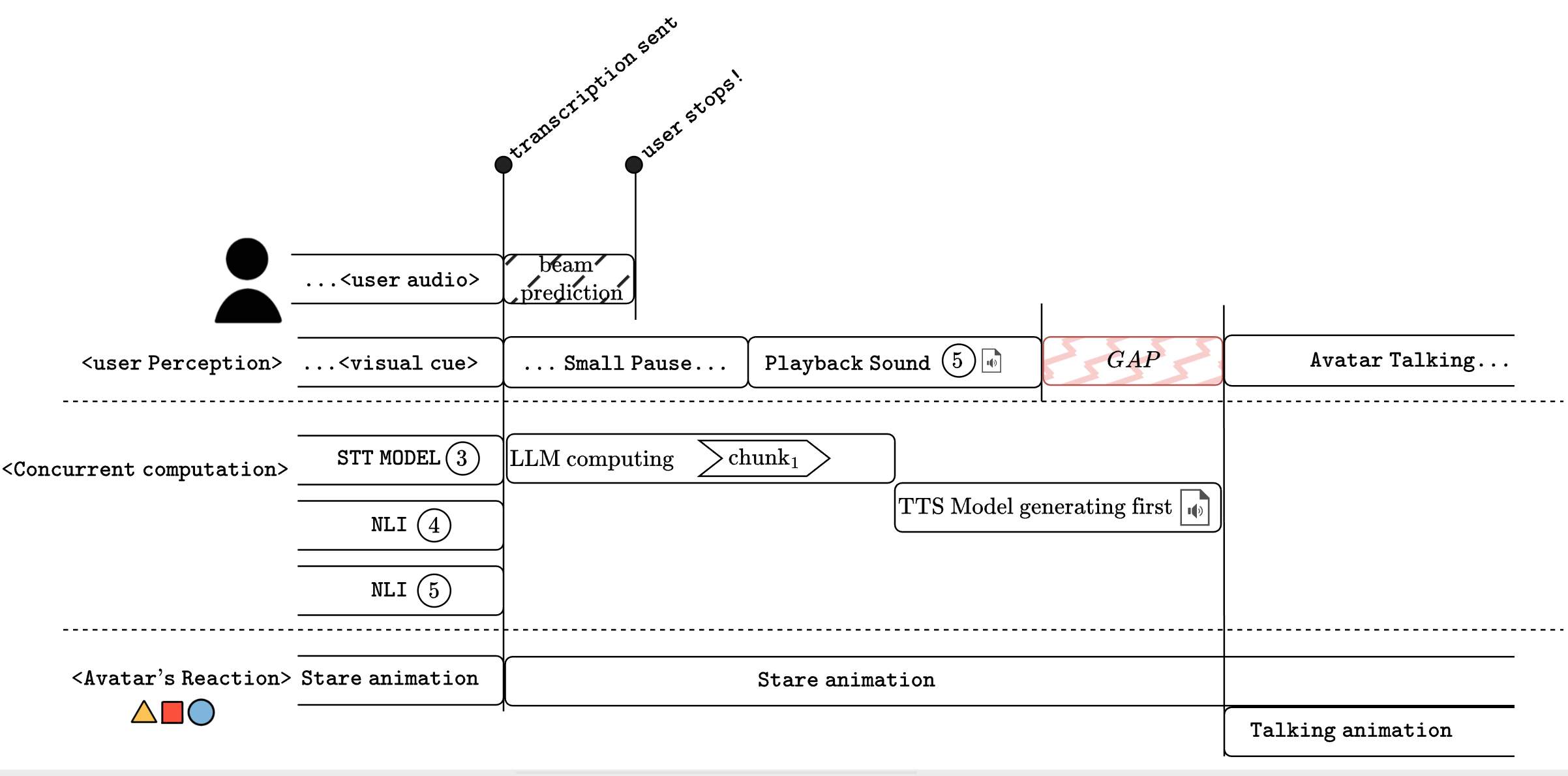}
    \caption{Temporal execution pipeline illustrating the \textit{Concurrency Problem}. Computational latency (STT, LLM, and TTS) is perceptually masked through immediate acoustic (Playback Sound) and visual (Stare animation) feedback before the generated response is ready.}
    \label{fig:concurrency_forms}
\end{figure*}

In natural human discourse, turn-taking involves instantaneous non-verbal cues. In voice-driven AI, the computational latency required for STT finalization, LLM generation, and initial Text-to-Speech (TTS) synthesis creates a silent interval, defined as the \textit{GAP} (Figure \ref{fig:concurrency_forms}). Human perception of conversational latency is highly sensitive, while users tolerate 300--600ms delays in telecommunications, fluid face-to-face dynamics require sub-60ms reactions, and delays exceeding 1000ms fundamentally break cognitive immersion.

The core objective of FORMS is to minimize the \textit{perceived} GAP to the strict $<60$ms threshold. Rather than eliminating the computational pipeline, FORMS masks it via a parallel Reaction NLI Ranker. This component evaluates the user's utterance against predefined emotional intents (e.g., thinking, surprised). Upon classification, it instantly retrieves a pre-computed acoustic filler specific to the targeted avatar's voice and triggers a visual gaze animation. By providing immediate multimodal feedback, the system satisfies the user's expectation of a response, neutralizing the perceived computational latency while the subsequent LLM generation occurs concurrently.

\subsection{Cognitive Inference and Presentation}
Once the active speaker is designated, the Cognitive Layer synthesizes the avatar's psychological profile, memory, and global context. This prompt is dispatched to low-latency LLM inference endpoints. To balance generation speed with rhetorical fidelity, the generated text is streamed in semantic chunks to an on-streaming TTS model. The TTS dynamically modulates prosody based on the initial emotional intent calculated by the Reaction NLI.

Finally, the Presentation Layer delivers the synchronized audio and visual feedback. Crucially, the avatars are abstracted into basic, highly distinguishable geometric shapes (e.g., a blue circle, a red square). This deliberate design choice suppresses visual anthropomorphic bias—particularly critical given the gendered perception of synthesized voices—ensuring that the socio-cognitive impacts measured in this study are driven exclusively by the semantic and interactive competence of the multi-agent system.

\section{Technical Evaluation}\label{sec:techeval}

To ensure the socio-cognitive effects measured in this study are attributable to the system's interactive quality rather than tolerated telecommunication delays, we benchmarked the FORMS architecture under real-time, human-intermediated execution. All local inference components (STT, NLI classification, and TTS) were executed on a single NVIDIA RTX 4080 Ti with 16GB of VRAM, demonstrating the framework's viability on consumer-grade hardware. It is crucial to note that the Speech-to-Text (STT) transcription delay is functionally negligible in our latency calculation. Because FORMS employs an aggressive real-time streaming approach utilizing \texttt{faster-whisper} with predictive beam search, words are transcribed and predicted ahead of the user's final pause. Consequently, STT execution overlaps with user speech and does not add a cumulative temporal penalty to the system's response readiness.

\subsection{Absolute Computational Latency ($L_c$)}
The Absolute Computational Latency ($L_c$) is defined as the temporal window from the user's final speech frame to the moment the system is ready to output synthesized audio. We evaluated two distinct TTS pipelines: the highly expressive baseline (\texttt{Qwen3-TTS} \cite{chu2023qwen}) and an ultra-low latency alternative (\texttt{Kokoro}). 

The measured latencies for the NLI (\texttt{mDeBERTa-v3} \cite{he2023debertav3}), the LLM Time-to-First-Token (\texttt{llama-3.1-8b-instant} \cite{grattafiori2024llama3herdmodels} via Groq), and the TTS engines across 100 initial interactions are detailed in Table \ref{tab:latency_metrics}.

\begin{table}[tb]
\centering
\caption{Computational Latency ($L_c$) by Component and Pipeline}
\label{tab:latency_metrics}
\renewcommand{\arraystretch}{1.2}
\begin{tabular}{lccc}
\hline
\textbf{Metric} & \textbf{Avg (ms)} & \textbf{p90 (ms)} & \textbf{p99 (ms)} \\ \hline
\multicolumn{4}{c}{\textit{Individual Components}} \\ \hline
NLI (mDeBERTa) & 18.44 & 22.09 & 25.07 \\
LLM TTFT (Groq) & 175.64 & 189.11 & 200.26 \\
TTS (Qwen) & 850.32 & 920.37 & 977.57 \\
TTS (Kokoro) & 32.91 & 41.69 & 48.86 \\ \hline
\multicolumn{4}{c}{\textit{Total Pipeline Latency ($L_c$)}} \\ \hline
\textbf{Qwen Base} & \textbf{1044.40} & \textbf{1115.67} & \textbf{1172.62} \\
\textbf{Kokoro Fast} & \textbf{226.99} & \textbf{243.02} & \textbf{256.25} \\ \hline
\end{tabular}
\end{table}

\subsection{Perceptual Masking and the Perceived GAP}
For the baseline Qwen architecture, the average $L_c \approx 1044.40$ ms creates an unacceptable conversational void. FORMS mitigates this via the NLI Reaction Engine, calculating the Perceived GAP ($GAP_p$) as:
$$ GAP_p = \max(0, L_c - D_{filler}) $$
Where $D_{filler}$ is the duration of the non-verbal acoustic reaction (e.g., a sigh, a laugh). As shown in Table \ref{tab:gap_comparison}, heavy cognitive states completely neutralize the computational latency ($GAP_p = 0$ ms). Conversely, shorter reactions leave a perceptible gap that is visually masked by the instantaneous avatar animations.

In contrast, for the ultra-low latency \texttt{Kokoro} pipeline, the model lacks the fine-grained expressiveness required to natively generate these complex non-verbal acoustic reactions. Consequently, there is no acoustic masking ($D_{filler} = 0$), meaning the perceived gap is directly equivalent to the system's total computational latency ($GAP_p = L_c$).

\begin{table}[tb]
\centering
\caption{Perceived GAP ($GAP_p$) by Intent (Baseline)}
\label{tab:gap_comparison}
\renewcommand{\arraystretch}{1.2}
\begin{tabular}{lc | ccc}
\hline
\textbf{Intent} & \textbf{$D_{filler}$} & \textbf{Avg} & \textbf{p90} & \textbf{p99} \\ \hline
Amused & 2090 ms & 0.00 & 0.00 & 0.00 \\
Thinking & 1548 ms & 0.00 & 0.00 & 0.00 \\
Disagreeing & 1492 ms & 0.00 & 0.00 & 0.00 \\
Realization & 984 ms & 64.23 & 131.67 & 188.62 \\
Sympathy & 956 ms & 89.74 & 159.67 & 216.62 \\
Disgusted & 932 ms & 112.90 & 183.67 & 240.62 \\
Neutral Filler & 930 ms & 114.86 & 185.67 & 242.62 \\
Surprised & 730 ms & 314.40 & 385.67 & 442.62 \\
Agreeing & 668 ms & 376.40 & 447.67 & 504.62 \\
Confused & 608 ms & 436.40 & 507.67 & 564.62 \\ \hline
\end{tabular}
\end{table}

\subsection{Architectural Evolution: Kokoro and the LLM Bottleneck}
Replacing the TTS component with \texttt{Kokoro} profoundly alters the pipeline's temporal constraints. Achieving an average stream latency of just $32.91$ ms, the total $L_c$ plummets to $226.99$ ms. Visually, this architectural shift is equivalent to completely removing the acoustic ``Playback Sound'' masking block from the temporal pipeline previously detailed in Figure \ref{fig:concurrency_forms}. 

Because $L_c$ natively approaches human cognitive thresholds, complex intent-specific soundbanks are no longer required to sustain the illusion of fluidity. By simply subtracting a brief $30$ ms artificial pacing pause, injected to prevent an unnatural, machine-like immediate overlap, the Kokoro architecture sustains a consistent, intent-agnostic $GAP_p$ of $196.99$ ms across all interactions. 

This extreme TTS optimization reveals a critical insight: the primary systemic bottleneck has entirely shifted to the LLM. In this configuration, the $175.64$ ms TTFT from Groq accounts for over 77\% of the system's absolute delay. 

To bypass this LLM bottleneck in future iterations, the FORMS architecture paves the way for a dual-LLM routing paradigm. In this approach, a highly optimized, smaller model generates the initial conversational chunk (the scaffolding) at extreme speeds. While the TTS streams this immediate audio to the user, the initial prompt and the generated first chunk are passed in parallel to a larger, high-capacity model (e.g., Llama 3 70B or GPT-4). The larger model then seamlessly takes over, completing the sentence and continuing the argumentative thread with deeper rhetorical fidelity and persona adherence. The structural and societal implications of this dual-routing approach are discussed further in Section \ref{sec:discussion}.

\section{System Test and Evaluation Methodology}\label{sec:experimentalmeth}

To assess the socio-cognitive impact of the FORMS architecture on human information processing, we conducted a system test and evaluation study. The study instantiated a moderated, multi-party conversational setting in which autonomous AI personas debated each other. A human moderator connected the synthetic panel to the audience, retaining exclusive authority to interrupt. The primary goal was to measure shifts in participants' trust calibration and cognitive security. Given the distinct scenarios, statistical analyses are conducted \emph{within-cohort}, avoiding direct cross-cohort numerical comparisons.

\subsection{Participants and Demographics}
We recruited $N=145$ participants, divided into a primary experimental cohort and a specialized pilot group to assess demographic vulnerabilities.
\begin{itemize}
    \item \textbf{Primary Cohort (Adolescents, $n=120$):} Students aged 14--17. Gender distribution: $51.7\%$ male, $43.3\%$ female, $5.0\%$ non-binary. Self-reported AI usage: occasional $59.2\%$, frequent $22.5\%$, daily $10.8\%$, other $7.5\%$.
    \item \textbf{Pilot Cohort (Adults/Specialists, $n=25$):} Professionals spanning ages 18--75+ attending a policy session. Gender: $52\%$ female, $48\%$ male. Education: PhD $44\%$, Master's $20\%$, other $34\%$. This group serves as an exploratory pilot study rather than an experimental control.
\end{itemize}

Procedures complied with institutional ethical standards. Informed consent/assent was secured for all participants. Data were anonymized and aggregated.

\subsection{Design, Scenarios, and Persona Configuration}
We employed a pre-test/post-test quasi-experimental design. Participants completed a baseline pre-questionnaire, observed a 25-minute live synthetic debate, engaged in a 20-minute moderated Q\&A, and completed a post-questionnaire. 

Scenarios were adapted per cohort while preserving the tripartite panel structure and ideological conflict: the Primary Cohort observed a workshop on \textit{AI and Disinformation}, while the Pilot Cohort observed a debate on \textit{EU AI Sovereignty}. 

\begin{figure}[h]
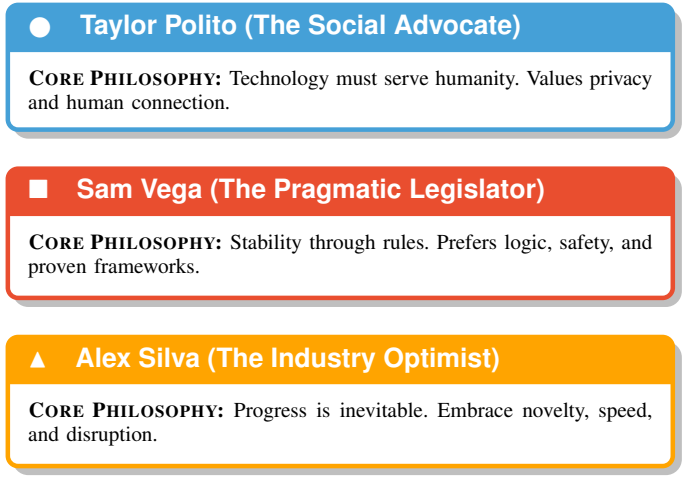

    \centering
    \begin{personacard}{$\CIRCLE$}{Taylor Polito (The Social Advocate)}{AndreaColor}
        \footnotesize \textbf{\textsc{Core Philosophy:}} Technology must serve humanity. Values privacy and human connection.
    \end{personacard}
    \vspace{1pt}
    \begin{personacard}{$\blacksquare$}{Sam Vega (The Pragmatic Legislator)}{SamColor}
        \footnotesize \textbf{\textsc{Core Philosophy:}} Stability through rules. Prefers logic, safety, and proven frameworks.
    \end{personacard}
    \vspace{1pt}
    \begin{personacard}{$\blacktriangle$}{Alex Silva (The Industry Optimist)}{AlexColor}
        \footnotesize \textbf{\textsc{Core Philosophy:}} Progress is inevitable. Embrace novelty, speed, and disruption.
    \end{personacard}
    \caption{Persona Configuration. Internal instructions adapt the role to the specific cohort scenario while preserving the underlying ideological conflict.}
    \label{fig:personas}
\end{figure}

During execution, the autonomous agents (Figure \ref{fig:personas}) were prompted to utilize semantic interpellation, directly addressing opposing arguments to sustain a coherent, high-fidelity argumentative thread. 

To isolate the persuasive impact of the synthetic rhetoric and mitigate human authority bias, the moderator adhered to a strict, neutral facilitation protocol. The moderator's interventions were exclusively restricted to procedural tasks: introducing the session, allocating the floor during the audience Q\&A, and enforcing time limits. Crucially, the moderator was forbidden from summarizing, validating, or reacting to the argumentative content generated by the agents, ensuring the measured socio-cognitive effects were driven by the autonomous interaction itself.

\subsection{Measures and Instruments}\label{subsec:measures}
Data were collected via paired pre- and post-intervention questionnaires utilizing a 5-point Likert scale ($1 =$ Strongly Disagree to $5 =$ Strongly Agree). Negatively phrased items were reverse-coded for consistency.

To isolate the psychological impact of the FORMS architecture and mitigate external confounding variables (e.g., prior media exposure), post-intervention items incorporated an explicit conditional preamble acting as a cognitive anchor (e.g., ``After observing the debate...''). Table \ref{tab:combined_measures} details the core measurement instruments across both cohorts, delineating variables designated for repeated-measures hypothesis testing (Comparable) versus post-only qualitative experience checks.

While the use of single-item measures precludes the calculation of standard internal consistency metrics (e.g., Cronbach's $\alpha$), we established the instrument's preliminary validity through two empirical methods on the baseline data. First, Spearman's rank correlation confirmed convergent validity between theoretically aligned constructs, such as baseline Confidence and perceived Reliability ($r_s = 0.512$, $p < 0.001$), alongside logical divergent validity against the demand for strict external regulation ($r_s = -0.188$, $p < 0.05$). Second, we confirmed known-group validity by segmenting the primary cohort by self-reported AI usage. A Welch's independent t-test revealed that the instrument successfully discriminated baseline confidence levels: high-frequency users (daily/often) reported significantly higher initial confidence ($M = 3.08$) compared to low-frequency users (sometimes/never) ($M = 2.34$, $p < 0.001$). Together, these tests confirm that the questionnaire captures distinct, sensitive dimensions of human-AI interaction, adequately serving the exploratory scope of this study.

\begin{table*}[t]
\centering
\caption{Measurement Instruments: Primary Cohort (Adolescents) and Pilot Cohort (Adults)}
\label{tab:combined_measures}
\renewcommand{\arraystretch}{1.1}
\scriptsize
\begin{tabular}{p{0.08\textwidth} p{0.75\textwidth} p{0.12\textwidth}}
\hline
\textbf{Metric Type} & \textbf{Item Formulation (Post-Intervention Phrasing or Anchor)} & \textbf{Pre/Post Paired} \\
\hline
\multicolumn{3}{c}{\textbf{Primary Cohort: Adolescents (Scenario A)}} \\
\hline
Confidence & \emph{After seeing the interaction, how would you rate your confidence to use AI for school tasks?} & Yes \\
Reliability & \emph{Seeing AIs debate suggests that these systems can provide reliable and well-reasoned information.} & Yes \\
Influence & \emph{After observing the debate, I believe AI's ability to influence what we think is...} & Yes \\
Sincerity & \emph{Compared to humans, how sincere did this debate between AIs seem to you?} & Yes \\
Regulation & \emph{Has your opinion changed about the need for external laws for AI?} & Yes \\
Strategy & \emph{After this experience, what should be the priority? [Multiple choice]} & Yes \\
Qualitative & \emph{Describe the experience in a single word. / Were you surprised it had ``opinions''?} & Post-only \\
\hline
\multicolumn{3}{c}{\textbf{Pilot Cohort: Adults/Specialists (Scenario B)}} \\
\hline
Comfort & \emph{Having interacted with GenAI, I would feel less comfortable relying on it in my professional work.} (Rev) & Yes \\
Trust & \emph{Having seen AI models ``debate'' themselves made me trust more in their ability to provide reliable info.} & Yes \\
Manipulation & \emph{Having seen AI models ``debate'', I believe GenAI is likely to manipulate beliefs at scale.} & Yes \\
Truthfulness & \emph{For me this debate between AI models was more truthful than similar debates with real persons.} & No (Post-only) \\
Protection & \emph{Having interacted with it, do you feel you need to be protected from Generative AI?} & Yes \\
Policy Role & \emph{Having seen AI models ``debate'' themselves, should policymakers priority be to: [Multiple choice]} & No (Post-only) \\
\hline
\end{tabular}
\end{table*}

\section{System Test and Evaluation Results}\label{sec:results}

Data analysis was performed across the primary experimental cohort (adolescents) and the specialized pilot cohort (adults) using appropriate non-parametric tests for categorical shifts (McNemar) and parametric tests (paired and independent t-tests) for continuous variables. To establish the magnitude of the observed phenomena, Cohen's $d$ effect sizes were calculated for all parametric tests. The alpha level for statistical significance was set at $0.05$. The results are structured around three core dimensions: governance shifts, trust calibration in human-machine interaction, and the psychological triggers of regulatory demand. Table \ref{tab:hypotheses} summarizes the updated hypotheses and statistical outcomes.

\begin{table*}[t]
\centering
\caption{Summary of Hypotheses and Statistical Outcomes}
\label{tab:hypotheses}
\renewcommand{\arraystretch}{1.2}
\scriptsize
\begin{tabular}{p{0.05\textwidth} p{0.35\textwidth} p{0.05\textwidth} p{0.35\textwidth}}
\hline
\textbf{ID} & \textbf{Description} & \textbf{Outcome} & \textbf{Key Evidence} \\
\hline
H1 & \textbf{Cognitive Destabilization:} Exposure to competent MAS debate fragments prior strategic governance preferences rather than polarizing users toward a single stance. & Supported & 38.4\% stance plasticity; 2x2 McNemar test showed symmetrical flow, no net shift to strict regulation ($p=1.000$). \\
H2 & \textbf{Regulatory Awakening:} Interaction increases the explicit baseline demand for external legal frameworks. & Supported & One-sample t-test ($M=3.48$, $p<.001$, $d=0.44$). \\
H3 & \textbf{Skepticism Induction:} Realistic exposure lowers blind trust in AI reliability for concrete tasks and recalibrates exaggerated fears of influence. & Supported & Trust drop ($p=.055$, $d=0.18$); lowered influence estimation among baseline skeptics ($p=.001$, $d=-0.48$). \\
H4 & \textbf{Surprise-Regulation Link:} Anthropomorphic surprise (AI holding "opinions") acts as a catalyst for demanding policy intervention. & Supported & Surprised subjects demanded stricter laws ($p=.031$, $d=0.43$). \\
H5 & \textbf{The Normality Paradox:} High-quality, fluid interactions trigger higher regulatory alarm than erratic or poor interactions. & Supported & ``Useful/Interesting" experiences yielded higher regulation demand ($M=3.57$) vs. ``Weird/Negative" ($M=2.88$, $p=.036$, $d=0.65$). \\
H6 & \textbf{Truthfulness Paradox:} Adult pilot cohort exhibits a cynicism gap, perceiving AI agents as more sincere in debate than human equivalents. & Supported & 91.3\% agreement; One-sample t-test ($M=4.17$, $p<.001$, $d=2.04$). \\
\hline
\end{tabular}
\end{table*}

\subsection{Governance Shift and Cognitive Destabilization (H1 \& H2)}
Initial assumptions in AI safety often posit that highly persuasive systems might polarize users toward a specific ideological or regulatory stance. Our data reveals a more complex phenomenon regarding human information processing: cognitive destabilization.

When evaluating the overarching strategic priority for AI governance (P6: e.g., Education vs. Freedom vs. Strict Limits), we observed a high degree of stance plasticity. Specifically, 38.4\% of the primary cohort changed their primary preferred approach after observing the debate. However, contrary to the initial hypothesis of a unidirectional shift toward regulation, a 2x2 McNemar test isolating the "Strict Limits" category revealed a perfectly symmetrical flow ($p=1.000$). The number of participants adopting a strict regulatory stance was offset by an equal number migrating away from it toward education or laissez-faire approaches.

Despite this fragmentation regarding \textit{how} to govern AI (H1), there was a robust consensus on the \textit{need} for governance (H2). When asked explicitly about the necessity of external legal frameworks (P5), the primary cohort demonstrated a significant ``Regulatory Awakening.'' A one-sample t-test against the neutral baseline confirmed a strong post-intervention demand for laws ($M=3.48, p<.001$), yielding a small but practically consistent effect size ($d=0.44$). Given that this shift occurred after a single 25-minute exposure, it underscores the immediate persuasive capacity of the architecture.

\begin{figure}[h]
    \centering
    \includegraphics[width=0.48\textwidth]{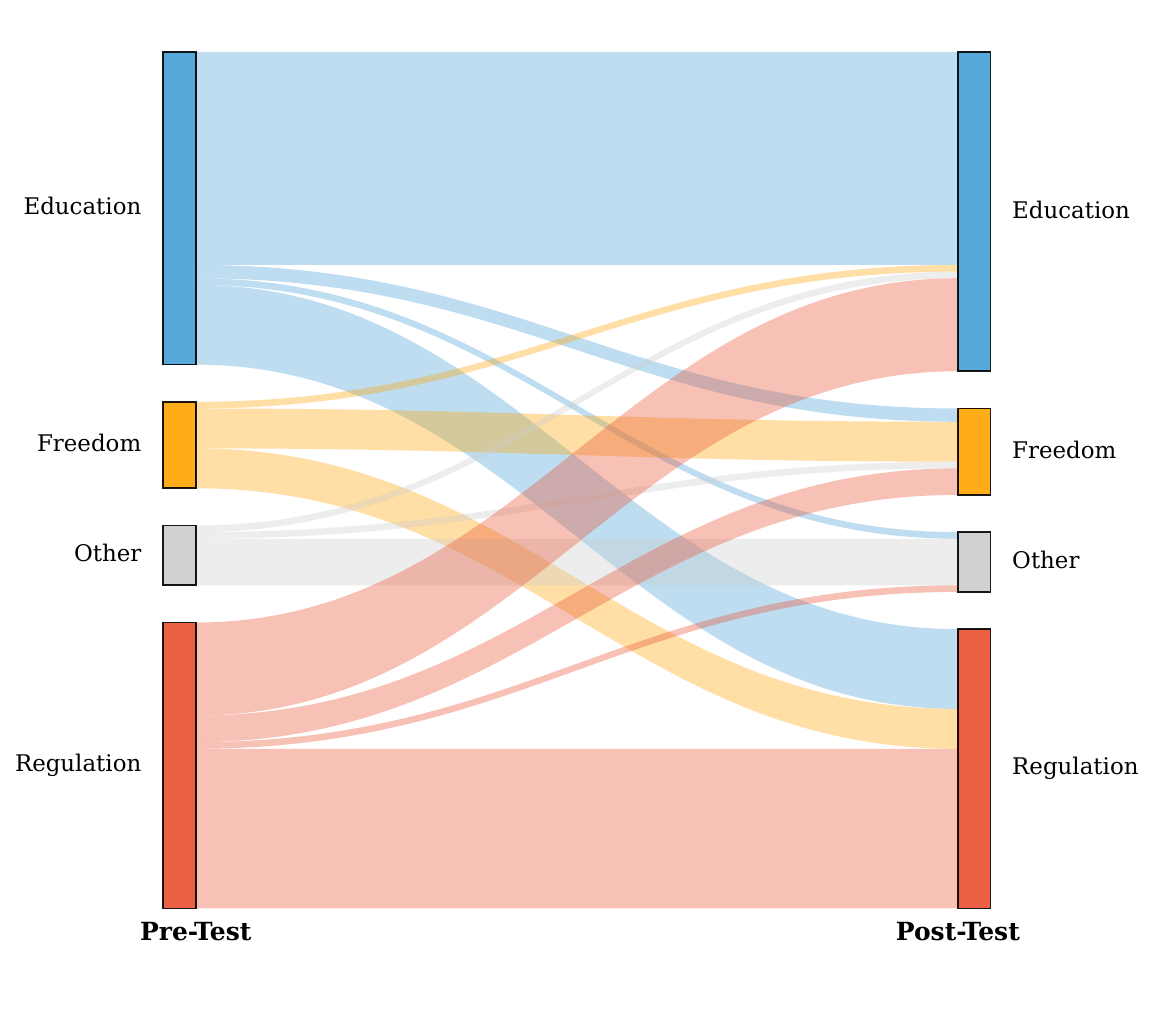}
    \caption{Flow of Governance Preferences (Pre vs. Post). The multidirectional migration illustrates cognitive destabilization, with colors corresponding to the rhetorical stances of the autonomous agents (Regulation/Red, Freedom/Yellow, Education/Blue).}
    \label{fig:sankey}
\end{figure}

\subsection{Trust Calibration and the Truthfulness Paradox (H3 \& H6)}
The architecture's low latency and conversational fluency produced divergent effects on trust calibration and information processing across demographics.

\subsubsection*{Primary Cohort: Skepticism Induction}
For the primary cohort, observing the MAS debate served as a calibrating event. We observed a decrease in blind trust regarding the reliability of AI for school tasks, though this shift was only marginally significant with a negligible effect size ($M_{pre}=2.90$ vs. $M_{post}=2.71, p=.055, d=0.18$). However, analyzing the subset of users who initially estimated AI's influence capacity as low (H3b) revealed a significant reduction in their estimation with a practical effect size ($p=.001, d=-0.48$). This nuanced outcome supports H3, suggesting that while deep-seated generalized trust is resilient, grounded exposure effectively recalibrates specific exaggerated perceptions of algorithmic capabilities, fostering a healthier skepticism.

\begin{figure}[h]
    \centering
    \includegraphics[width=0.45\textwidth]{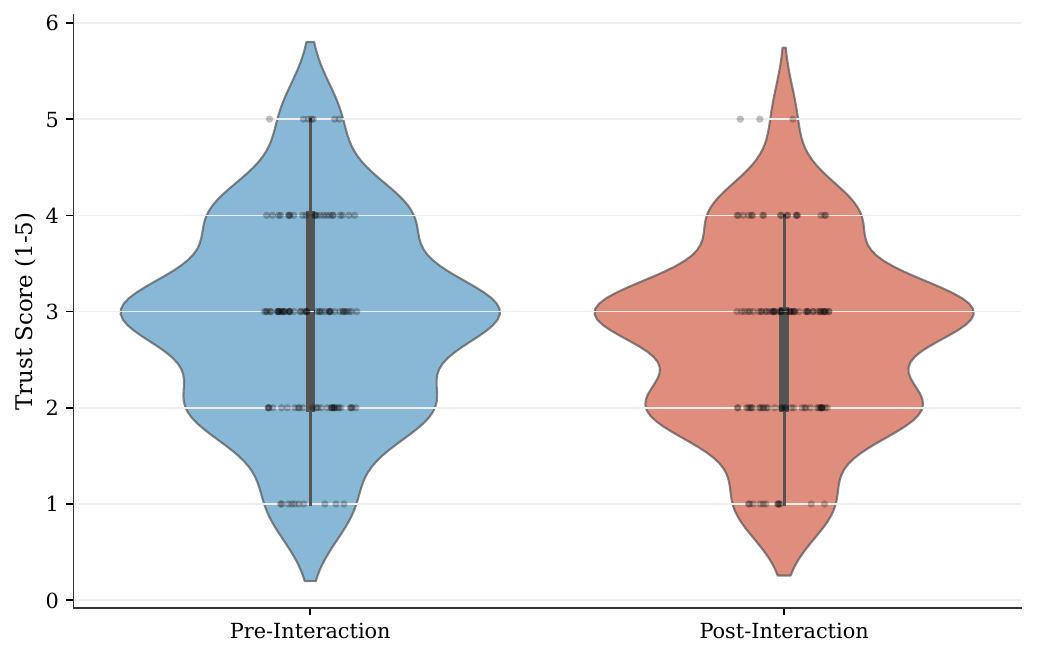}
    \caption{Distribution shift in primary cohort trust regarding AI reliability for school tasks. The flattening of the distribution curve indicates skepticism induction post-interaction.}
    \label{fig:trust_shift}
\end{figure}

\subsubsection*{Pilot Cohort: The Truthfulness Paradox}
Conversely, the adult pilot cohort exhibited a critical cognitive vulnerability. When evaluating the sincerity of the synthetic debate compared to human political equivalents, 91.3\% of the pilot cohort agreed that the AI interaction was more truthful, yielding a highly significant outcome ($M=4.17/5.00, p<.001$). Crucially, the analysis revealed a highly pronounced effect size ($d=2.04$). While requiring careful interpretation due to the constrained sample size of the pilot group, this magnitude strongly supports H6 (The Truthfulness Paradox). It provides preliminary evidence that, although professional adults recognize the risks of generative systems, their existing cynicism toward human political discourse may lead them to attribute a false equivalence of objectivity and sincerity to the neutral tone of an LLM-driven architecture. This phenomenon is deeply rooted in the ``Machine Heuristic,'' a cognitive bias where individuals inherently attribute greater objectivity, lower bias, and independent effort to algorithmic outputs compared to human equivalents, effectively bypassing critical evaluation filters \cite{a872d9a42e5b41b99ef69f4257521f4c, HEIMSTAD2025100190}.

\begin{figure}[h]
    \centering
    \includegraphics[width=0.48\textwidth]{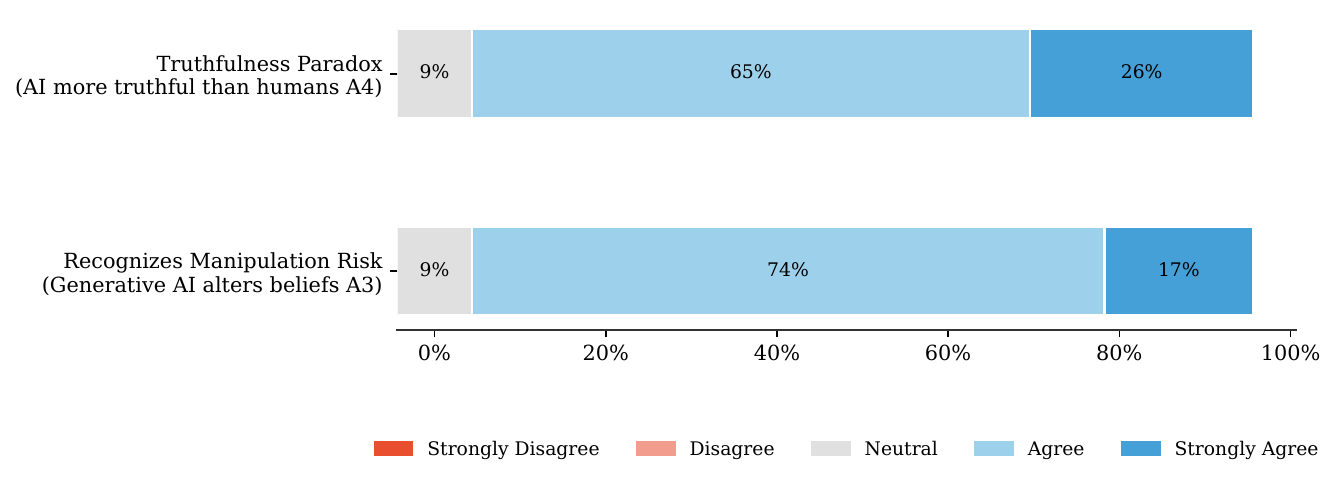}
    \caption{The Truthfulness Paradox. A diverging stacked bar chart highlighting the cognitive dissonance in the pilot cohort: simultaneous recognition of manipulation risk and assignment of high truthfulness to synthetic entities.}
    \label{fig:truthfulness}
\end{figure}

\subsection{Cognitive Ergonomics and the Triggers of Regulation (H4 \& H5)}
To understand the mechanisms driving the regulatory demand observed in H2, we analyzed subjective user experiences.

\subsubsection*{The Surprise-Regulation Link (H4)} The manifestation of synthetic ``opinions" provoked a defensive response. Participants who expressed surprise that an AI could sustain an ideological stance demanded significantly stricter legal interventions ($p=.031, d=0.43$) compared to their non-surprised peers. 

\subsubsection*{The Normality Paradox (H5)} Traditional AI safety literature often associates system failures (e.g., hallucinations, erratic outputs) with public mistrust. Our framework demonstrates the inverse. The open-ended single-word responses describing the experience (F1) were semantically coded into two primary valences by independent researchers: fluid/positive interactions (e.g., ``Useful,'' ``Interesting'') and erratic/negative interactions (e.g., ``Weird,'' ``Boring''). Participants who categorized the interaction as ergonomically fluid demanded significantly more regulation ($M=3.57$) than those who perceived the interaction as erratic ($M=2.88$). An independent t-test confirmed this difference ($p=.036$), revealing a medium effect size ($d=0.65$). This robust effect quantifies the ``Normality Paradox'': it indicates that societal alarm is triggered not by algorithmic failure, but by algorithmic competence and its seamless integration into human conversational norms.

\subsection{Limitations, Moderator Bias, and Statistical Robustness}
While the findings present strong theoretical implications, we must acknowledge the presence of a human moderator as a confounding variable in the experimental design. The moderator was introduced to emulate real-world panel dynamics and manage the flow of the debate. However, human authority validating the synthetic interaction might inherently elevate the perceived legitimacy and sincerity of the agents. Future experimental designs must evaluate entirely unmoderated human-machine environments to isolate the baseline persuasive power of the synthetic architecture.

Furthermore, the sample size of the adult pilot cohort ($n=25$) restricts the findings to an exploratory, albeit critical, status. In traditional social sciences, such a constrained cohort requires cautious extrapolation. Nevertheless, the calculation of Cohen's $d$ yielded a massive effect size ($d=2.04$) regarding the Truthfulness Paradox. Statistically, an effect size of this magnitude in a highly educated demographic suggests that the cognitive vulnerability (the signal) overwhelmingly breaches standard critical heuristics (the noise). Rather than a conclusive population-level generalization, this pilot serves as a successful cognitive stress test. It strongly indicates that theoretical AI literacy does not inherently prevent susceptibility to the Machine Heuristic, justifying the immediate need for larger, cross-cultural replication studies to establish population-level baselines.

Finally, we acknowledge that the measurement instrument utilized single-item, ad-hoc Likert scales rather than standardized psychometric batteries. While this choice deliberately minimized cognitive fatigue in the adolescent cohort immediately following a demanding cognitive task, it limits exhaustive test-retest reliability assessments. Future iterations will translate these preliminary single-item findings into validated, multi-item psychometric frameworks.

\section{Discussion and Future Work}\label{sec:discussion}

The central premise of this paper is that the psychological impact of multi-agent AI systems is inextricably linked to their technical and rhetorical competence. Our empirical evaluation demonstrates that as these systems cross critical thresholds of conversational fluidity, they trigger unique cognitive vulnerabilities that traditional AI safety paradigms, which largely focus on algorithmic failure or hallucination, fail to capture.

\subsection{Synthesizing Competence and Cognitive Vulnerability}
The magnitude of the observed statistical effects provides critical insights for future Human-AI Interaction (HAI) frameworks. The medium effect size ($d=0.65$) associated with the \textit{Normality Paradox} confirms that interaction fluidity directly scales with societal regulatory alarm. More alarmingly, the massive effect size ($d=2.04$) observed in the \textit{Truthfulness Paradox} among specialized adults indicates a critical blind spot in current cognitive security models. It demonstrates that theoretical AI literacy, inherent in professional cohorts who understand that models can hallucinate or manipulate, does not inoculate users against the ``Machine Heuristic'' when exposed to high-fidelity, autonomous multi-agent debates. The neutral, objective cadence orchestrated by the MAS effectively bypasses critical filters. These empirical baselines prove that persuasive risks are immediate consequences of deploying low-latency MAS, justifying the need to scrutinize the underlying architectural pipelines.

\subsection{Architectural Roadmap: Resolving Network Dependency via Dual-Routing}

A valid critique of relying on external LLM inference endpoints is the inherent dependency on network stability, which can introduce stochastic jitter and undermine the deterministic latency required for fluid interaction. To decouple the architecture from cloud reliance and eliminate the LLM bottleneck, future iterations of FORMS will implement a \textit{Hybrid Dual-Routing LLM Architecture}.

In this paradigm, the orchestration layer routes the finalized user prompt to two distinct models to balance extreme speed with rhetorical depth:
\begin{enumerate}
    \item \textbf{The Local Scaffolding Model:} A highly optimized, small conversational model running entirely on local edge hardware. This model is tasked exclusively with instantly generating a valid, context-aware conversational opening (e.g., \textit{"That is a valid concern, however..."}). Because it executes locally, it guarantees a deterministic, zero-network-dependency Time-to-First-Token (TTFT).
    \item \textbf{The High-Capacity Reasoning Model:} Concurrently, the same prompt is sent to a larger, high-capacity model (either via a fast API or a larger local instance). While the local scaffolding audio is streaming to the user, this secondary model seamlessly takes over to complete the sentence and sustain the core argumentative payload.
\end{enumerate}

This hybrid approach ensures that the critical sub-60ms reaction threshold is met securely on local hardware, while leveraging more powerful models for rhetorical fidelity, effectively neutralizing API latency fluctuations. Furthermore, for deployments requiring strict data sovereignty or air-gapped security, this dual-routing approach can be implemented entirely locally by pairing a small local model for scaffolding with a moderately sized local model for continuation. This proves a critical point: highly persuasive, real-time MAS interactions can be achieved entirely on consumer-grade hardware without relying on extreme computational infrastructure.

\subsection{Roadmap and Empirical Validation}
To ensure full reproducibility, all tested configurations, generation parameters, and system prompts used in this study are currently available in the open-source repository of the \href{https://github.com/marcosrv-ULL/FORMS/tree/main}{FORMS framework}. Furthermore, upon the formal acceptance of this paper, the proposed dual-routing architecture will also be integrated and publicly released within this repository.

Our immediate future work entails deploying this zero-filler, high-cognition architecture in a new experimental wave. By testing a larger and more demographically diverse participant sample, we aim to measure whether an interaction completely devoid of latency-masking artifacts further amplifies the \textit{Normality Paradox}. We hypothesize that the total elimination of conversational friction will increase the rhetorical persuasion of the agents, thereby reinforcing the observed socio-cognitive shifts and driving an even higher demand for regulatory intervention.

\section{Conclusion}\label{sec:conclusion}
As generative AI systems transition from sequential text interfaces to spatially mediated, multi-agent conversational environments, the vectors of psychological influence evolve. This paper introduced FORMS, an open-source architecture that solves the concurrency problem in multi-agent voice interactions, and utilized it to empirically measure the socio-cognitive impact of autonomous synthetic debates.

Our findings reveal that exposure to competent, multi-agent AI debates triggers \textit{Cognitive Destabilization}, fragmenting users' prior strategic consensus by presenting conflicting viewpoints with equal rhetorical validity. Concurrently, we observed a robust \textit{Regulatory Awakening} across demographics. Importantly, this societal alarm is not driven by algorithmic failure, but by algorithmic competence, a phenomenon we define and quantify as the \textit{Normality Paradox}, where the seamless integration of AI into human conversational norms acts as the primary catalyst for demanding external legal frameworks. 

Finally, the preliminary identification of the \textit{Truthfulness Paradox} within our specialist cohort underscores a potentially critical societal vulnerability: even among demographics that theoretically recognize the risks of generative AI, there is a pronounced tendency to assign greater sincerity to a synthetic LLM debate than to equivalent human discourse. This indicates that the strict, probabilistic turn-taking orchestration, which successfully masks computational latency and mimics human conversational rhythm, effectively bypasses critical evaluation filters. Supported by substantial statistical effect sizes, these findings confirm that as the hardware and architectural barriers to real-time, highly expressive MAS continue to fall, ensuring cognitive security will require regulatory frameworks that evaluate not just the factual accuracy of what these models know, but the profound psychological impact of how persuasively they can deliver it.

\section*{Acknowledgments}
This work has been possible thanks to the Binter Cybersecurity Chair at the University of La Laguna, and to the research projects PID2022-138933OB-I00 (ATQUE) and 2023DIG28 (IACTA), funded by MCIN/AEI/10.13039/501100011033/FEDER, EU, and the CajaCanarias Foundation / "la Caixa", respectively.

\bibliographystyle{ieeetr}
\bibliography{bibliography}

@misc{labs2025mercuryultrafastlanguagemodels,
      title={Mercury: Ultra-Fast Language Models Based on Diffusion}, 
      author={Inception Labs and Samar Khanna and Siddhant Kharbanda and Shufan Li and Harshit Varma and Eric Wang and Sawyer Birnbaum and Ziyang Luo and Yanis Miraoui and Akash Palrecha and Stefano Ermon and Aditya Grover and Volodymyr Kuleshov},
      year={2025},
      eprint={2506.17298},
      archivePrefix={arXiv},
      primaryClass={cs.CL},
      url={https://arxiv.org/abs/2506.17298}, 
}

@misc{moon2024lpulatencyoptimizedhighlyscalable,
      title={LPU: A Latency-Optimized and Highly Scalable Processor for Large Language Model Inference}, 
      author={Seungjae Moon and Jung-Hoon Kim and Junsoo Kim and Seongmin Hong and Junseo Cha and Minsu Kim and Sukbin Lim and Gyubin Choi and Dongjin Seo and Jongho Kim and Hunjong Lee and Hyunjun Park and Ryeowook Ko and Soongyu Choi and Jongse Park and Jinwon Lee and Joo-Young Kim},
      year={2024},
      eprint={2408.07326},
      archivePrefix={arXiv},
      primaryClass={cs.AR},
      url={https://arxiv.org/abs/2408.07326}, 
}

@misc{hexgrad_2025,
	author       = { Hexgrad },
	title        = { Kokoro-82M (Revision d8b4fc7) },
	year         = 2025,
	url          = { https://huggingface.co/hexgrad/Kokoro-82M },
	doi          = { 10.57967/hf/4329 },
	publisher    = { Hugging Face }
}

@misc{maimon2025scalinganalysisinterleavedspeechtext,
      title={Scaling Analysis of Interleaved Speech-Text Language Models}, 
      author={Gallil Maimon and Michael Hassid and Amit Roth and Yossi Adi},
      year={2025},
      eprint={2504.02398},
      archivePrefix={arXiv},
      primaryClass={cs.CL},
      url={https://arxiv.org/abs/2504.02398}, 
}

@inproceedings{
chuang2025debate,
title={{DEBATE}: A Large-Scale Benchmark for Role-Playing {LLM} Agents in Multi-Agent, Long-Form Debates},
author={Yun-Shiuan Chuang and Ruixuan Tu and Chengtao Dai and Smit Vasani and Binwei Yao and Michael Henry Tessler and Sijia Yang and Dhavan V. Shah and Robert D. Hawkins and Junjie Hu and Timothy T. Rogers},
booktitle={Workshop on Scaling Environments for Agents},
year={2025},
url={https://openreview.net/forum?id=7mWVbd4IXD}
}

@misc{zhang2026verifiedmultiagentorchestrationplanexecuteverifyreplan,
      title={Verified Multi-Agent Orchestration: A Plan-Execute-Verify-Replan Framework for Complex Query Resolution}, 
      author={Xing Zhang and Yanwei Cui and Guanghui Wang and Wei Qiu and Ziyuan Li and Fangwei Han and Yajing Huang and Hengzhi Qiu and Bing Zhu and Peiyang He},
      year={2026},
      eprint={2603.11445},
      archivePrefix={arXiv},
      primaryClass={cs.AI},
      url={https://arxiv.org/abs/2603.11445}, 
}

@inproceedings{
choi2025debate,
title={Debate or Vote: Which Yields Better Decisions in Multi-Agent Large Language Models?},
author={Hyeong Kyu Choi and Jerry Zhu and Sharon Li},
booktitle={The Thirty-ninth Annual Conference on Neural Information Processing Systems},
year={2025},
url={https://openreview.net/forum?id=iUjGNJzrF1}
}

@misc{li2026dontbelievereadunderstanding,
      title={Don't believe everything you read: Understanding and Measuring MCP Behavior under Misleading Tool Descriptions}, 
      author={Zhihao Li and Boyang Ma and Xuelong Dai and Minghui Xu and Yue Zhang and Biwei Yan and Kun Li},
      year={2026},
      eprint={2602.03580},
      archivePrefix={arXiv},
      primaryClass={cs.CR},
      url={https://arxiv.org/abs/2602.03580}, 
}

@article{Weerts_2025, title={Generative AI in public administration in light of the regulatory awakening in the US and EU}, volume={1}, DOI={10.1017/cfl.2024.10}, journal={Cambridge Forum on AI: Law and Governance}, author={Weerts, Sophie}, year={2025}, pages={e3}}

@misc{chen2025turntakingsynchronousdialoguesurvey,
      title={From Turn-Taking to Synchronous Dialogue: A Survey of Full-Duplex Spoken Language Models}, 
      author={Yuxuan Chen and Haoyuan Yu},
      year={2025},
      eprint={2509.14515},
      archivePrefix={arXiv},
      primaryClass={cs.CL},
      url={https://arxiv.org/abs/2509.14515}, 
}

@article{HEIMSTAD2025100190,
title = {Machine heuristic in algorithm aversion: Perceived creativity and effort of output created by or with artificial intelligence},
journal = {Computers in Human Behavior: Artificial Humans},
volume = {5},
pages = {100190},
year = {2025},
issn = {2949-8821},
doi = {https://doi.org/10.1016/j.chbah.2025.100190},
url = {https://www.sciencedirect.com/science/article/pii/S294988212500074X},
author = {Sigurd Birk Heimstad and Anders Hauge Wien and Tarje Gaustad}
}

@article{NOVELLI2024106066,
title = {Generative AI in EU law: Liability, privacy, intellectual property, and cybersecurity},
journal = {Computer Law \& Security Review},
volume = {55},
pages = {106066},
year = {2024},
issn = {2212-473X},
doi = {https://doi.org/10.1016/j.clsr.2024.106066},
url = {https://www.sciencedirect.com/science/article/pii/S0267364924001328},
author = {Claudio Novelli and Federico Casolari and Philipp Hacker and Giorgio Spedicato and Luciano Floridi}
}

@article{Bai2025,
  author = {Bai, Hui and Voelkel, Jan G. and Muldowney, Shane and Eichstaedt, Johannes C. and Willer, Robb},
  title = {LLM-generated messages can persuade humans on policy issues},
  journal = {Nature Communications},
  volume = {16},
  number = {1},
  pages = {6037},
  year = {2025},
  doi = {10.1038/s41467-025-61345-5},
  url = {https://doi.org/10.1038/s41467-025-61345-5}
}

@article{Holbling2025,
  author = {H\"{o}lbling, Lukas and Maier, Sebastian and Feuerriegel, Stefan},
  title = {A meta-analysis of the persuasive power of large language models},
  journal = {Scientific Reports},
  volume = {15},
  number = {1},
  pages = {43818},
  year = {2025},
  doi = {10.1038/s41598-025-30783-y},
  url = {https://doi.org/10.1038/s41598-025-30783-y}
}

@misc{liu2026breakingmartingalecursemultiagent,
      title={Breaking the Martingale Curse: Multi-Agent Debate via Asymmetric Cognitive Potential Energy}, 
      author={Yuhan Liu and Juntian Zhang and Yichen Wu and Martin Takac and Salem Lahlou and Xiuying Chen and Nils Lukas},
      year={2026},
      eprint={2603.06801},
      archivePrefix={arXiv},
      primaryClass={cs.AI},
      url={https://arxiv.org/abs/2603.06801}, 
}

@misc{campedelli2025iwantbreakfree,
      title={I Want to Break Free! Persuasion and Anti-Social Behavior of LLMs in Multi-Agent Settings with Social Hierarchy}, 
      author={Gian Maria Campedelli and Nicolò Penzo and Massimo Stefan and Roberto Dessì and Marco Guerini and Bruno Lepri and Jacopo Staiano},
      year={2025},
      eprint={2410.07109},
      archivePrefix={arXiv},
      primaryClass={cs.CL},
      url={https://arxiv.org/abs/2410.07109}, 
}

@misc{rogiers2024persuasionlargelanguagemodels,
      title={Persuasion with Large Language Models: a Survey}, 
      author={Alexander Rogiers and Sander Noels and Maarten Buyl and Tijl De Bie},
      year={2024},
      eprint={2411.06837},
      archivePrefix={arXiv},
      primaryClass={cs.CL},
      url={https://arxiv.org/abs/2411.06837}, 
}

@misc{liu2025llmdangerouspersuaderempirical,
      title={LLM Can be a Dangerous Persuader: Empirical Study of Persuasion Safety in Large Language Models}, 
      author={Minqian Liu and Zhiyang Xu and Xinyi Zhang and Heajun An and Sarvech Qadir and Qi Zhang and Pamela J. Wisniewski and Jin-Hee Cho and Sang Won Lee and Ruoxi Jia and Lifu Huang},
      year={2025},
      eprint={2504.10430},
      archivePrefix={arXiv},
      primaryClass={cs.CL},
      url={https://arxiv.org/abs/2504.10430}, 
}

@inproceedings{ki-etal-2025-multiple,
    title = "Multiple {LLM} Agents Debate for Equitable Cultural Alignment",
    author = "Ki, Dayeon  and
      Rudinger, Rachel  and
      Zhou, Tianyi  and
      Carpuat, Marine",
    editor = "Che, Wanxiang  and
      Nabende, Joyce  and
      Shutova, Ekaterina  and
      Pilehvar, Mohammad Taher",
    booktitle = "Proceedings of the 63rd Annual Meeting of the Association for Computational Linguistics (Volume 1: Long Papers)",
    month = jul,
    year = "2025",
    address = "Vienna, Austria",
    publisher = "Association for Computational Linguistics",
    url = "https://aclanthology.org/2025.acl-long.1210/",
    doi = "10.18653/v1/2025.acl-long.1210",
    pages = "24841--24877",
    ISBN = "979-8-89176-251-0"
}

@inproceedings{a872d9a42e5b41b99ef69f4257521f4c,
title = "Machine heuristic: When we trust computers more than humans with our personal information",
author = "\{Shyam Sundar\}, S. and Jinyoung Kim",
note = "Publisher Copyright: {\textcopyright} 2019 Association for Computing Machinery.; 2019 CHI Conference on Human Factors in Computing Systems, CHI 2019 ; Conference date: 04-05-2019 Through 09-05-2019",
year = "2019",
month = may,
day = "2",
doi = "10.1145/3290605.3300768",
language = "English (US)",
series = "Conference on Human Factors in Computing Systems - Proceedings",
publisher = "Association for Computing Machinery",
booktitle = "CHI 2019 - Proceedings of the 2019 CHI Conference on Human Factors in Computing Systems",

}

@misc{chu2023qwen,
      title={Qwen-Audio: Advancing Universal Audio Understanding via Unified Large-Scale Audio-Language Models}, 
      author={Yunfei Chu and Jin Xu and Xiaohuan Zhou and Qian Yang and Shiliang Zhang and Zhijie Yan and Chang Zhou and Jingren Zhou},
      year={2023},
      eprint={2311.07919},
      archivePrefix={arXiv},
      primaryClass={eess.AS},
      url={https://arxiv.org/abs/2311.07919}, 
}

@misc{he2023debertav3,
      title={DeBERTaV3: Improving DeBERTa using ELECTRA-Style Pre-Training with Gradient-Disentangled Embedding Sharing}, 
      author={Pengcheng He and Jianfeng Gao and Weizhu Chen},
      year={2023},
      eprint={2111.09543},
      archivePrefix={arXiv},
      primaryClass={cs.CL},
      url={https://arxiv.org/abs/2111.09543}, 
}

@article{salvi2024conversational,
   title={On the conversational persuasiveness of GPT-4},
   volume={9},
   ISSN={2397-3374},
   url={http://dx.doi.org/10.1038/s41562-025-02194-6},
   DOI={10.1038/s41562-025-02194-6},
   number={8},
   journal={Nature Human Behaviour},
   publisher={Springer Science and Business Media LLC},
   author={Salvi, Francesco and Horta Ribeiro, Manoel and Gallotti, Riccardo and West, Robert},
   year={2025},
   month=may, pages={1645–1653} }

@misc{grattafiori2024llama3herdmodels,
      title={The Llama 3 Herd of Models}, 
      author={Aaron Grattafiori and others},
      year={2024},
      eprint={2407.21783},
      archivePrefix={arXiv},
      primaryClass={cs.AI},
      url={https://arxiv.org/abs/2407.21783}, 
}

\end{document}